# EuSpRIG TEAM work:

# Tools, Education, Audit, Management

David Chadwick (EuSpRIG Chair),
University of Greenwich, Greenwich, London SE10 9LS
D.R.Chadwick@gre.ac.uk

## 1. The T.E.A.M Approach

*Research on spreadsheet errors began over fifteen years ago. During that time, there has been ample evidence demonstrating that spreadsheet errors are common and nontrivial. Quite simply, spreadsheet error rates are comparable to error rates in other human cognitive activities and are caused by fundamental limitations in human cognition, not mere sloppiness. Nor does ordinary "being careful" eliminate errors or reduce them to acceptable levels.* Panko R [1]

In many ways, Ray has summed up the entire problem of spreadsheet errors - they are due to our fundamental human weaknesses, our limitations and inabilities. As Ray so rightly points out, just 'being careful' is not going to be enough to solve the problem. It is heartening, therefore, to see that EuSpRIG has attempted to address those main areas where our human weaknesses may be identified, accommodated, overcome and hopefully reduced to negligible effect. A review of all the papers from past EuSpRIG conferences shows that they may be seen to fall into four distinct areas: tools, education, auditing and management. Interestingly, these four areas form the indicative acronym TEAM.

- Tools: methodologies and software for spreadsheet building
  Rajalingham K. et al [5] [14], Chadwick et al [10], O'Beirne [12],
  Paine J. [13], Raffensperger J. [15]. Grossman T. [21], Knight D [17]

- Education: awareness-raising, teaching, research
  Rajalingham K et al [2], Ayalew Y et al [4], Cleary P M [8], Banks D. et al [20],
  Chadwick D [9][19][23], Panko R. [1], Butler R [6],

- Audit: methods and software tools,
  Nixon D [18], Hock C. et al [7], Ettema H et al [16], Clermont M et al [22],
  Croll G [25]

- Management : standards and controls for managing developments
  Butler R [11] [24], Hawker A [3], Chadwick D [19]

TEAM is indicative because team-work is possibly the key to solving the corporate spreadsheet problem involving, as it might, the use of development tools where they exist, education in error awareness at all levels and for common models, properly organised and applied audit procedures, and lastly, the adoption of sound management policies to support standards and controls.

*Sound development methods, standards and user education are the obvious preventive controls. ... The many examples good development practice is rarely codified into business procedures, and even when it is, the rules and restrictions it requires are not followed to any significant degree.* Butler R [6]

## 2. TOOLS

This is possibly the area where there is much scope for effect. There is, without doubt a need for a structured approach to model building with the aid of software tools to direct the spreadsheet developer, advice and/or prevent possible error situations, and document their actions. Work in this area so far presented to EuSpRIG has included approaches that require the use of programming languages to 'code' the logic of the model and then to construct the spreadsheet interface of rows and columns along with it's included logic. An example of this is the tool ModelMaster (MM).

*"MM was developed from a branch of mathematics known as category theory ... briefly stated, category theory, like logic, is a tool for studying mathematical and computational concepts, one concerned much more with their form than with their content."* Paine J [13]

Additionally, many researchers have cited the need for methods based around software engineering principles.

*"The application of software engineering principles to spreadsheets - spreadsheet engineering - has the potential to increase the productivity of spreadsheet programmers, decrease the frequency and severity of spreadsheet errors, enhance spreadsheet maintainability over time, and actually be implemented by spreadsheet users."* Grossman T. [21]

Rajalingham, Chadwick and Knight have done some work in this vein creating a tool that forms Jackson type structures to give a graph type description of a spreadsheet during development for easy checking of logic and which also permits reverse engineering of a spreadsheet into its underlying Jackson structure. Hence each spreadsheet model can be defined by its underlying canonical model which would remain intact even if the form of the spreadsheet were changed cosmetically.

*"Based on software engineering principles ... it has been found that spreadsheet models can be represented in a form identical to the data structure diagram developed by Jackson.... The proposed methodology demonstrates how these techniques can in fact be transferred to the production of spreadsheets."* Rajalingham et al [14]

Perhaps one of the more exciting and commercially useful tools has been the Brixx product, essentially a spreadsheet alternative.

*"The combination of the latest object technology combined with a new approach offers a genuine alternative to spreadsheets that goes some way towards satisfying the identified resistance. While challenges still remain, the future could hold an exciting new power and capability for all business modellers to meet the more complex modelling challenges of the future."* Knight D. [17]

Although not in existence, there might be some use for a tool to prompt users into properly documenting their spreadsheets.

*"Worksheets often duplicate some information that is also recorded elsewhere and was copied in by typing, copy & paste, programming code, database add-ins, and macros. However, modifying the original data source will usually not update the same information in the worksheet. Therefore, there is a high risk of creating inconsistencies between worksheets and other information systems. All this is not immediately obvious on reading a spreadsheet as they are rarely documented"* O'Beirne [12]

## 3. EDUCATION

Spreadsheet builders of all types and at all levels of an organisation require the necessary skills either through training, formal education or in-house seminars. Such training should encompass two types of skills: the possibly complicated but straight forward practical skills of manipulating a given software package (package functionality) and the more sophisticated and more subtle skills of comprehending the business requirement and forming a correct conceptual model of how it needs to be structured and what needs to be shown (modelling).

*"Nevertheless, nobody can and will write a spreadsheet without having such a conceptual model in mind, be it of numeric nature or a layout focussed, geometrical nature"* Ayalew et al [2]

However, there is some suggestion, and field work is presently under way to clarify this, that many spreadsheet practitioners are self-taught or receive only rudimentary training and that which they do receive is based solely upon package functionality, the first of the skills outlined above. There is some suggestion, though more work needs to be done in this area to determine the true scale of the problem that such training rarely touches upon the modelling process as outlined in the second of the skills above.

*"From a look at the training and teaching curricula of academia and higher education, there appears to be no attempt to present students with working standards for spreadsheet developments nor even an acknowledgement that such may be required"* Chadwick D. [19]

The weaknesses of formal education courses (see Chadwick D [19]) are often echoed in professional spreadsheet training courses by private consultants and in-house training seminars. This is a shame as business practitioners who are the main recipients of this training are being denied opportunities to become aware of the nature and extent of errors and that models benefit from some kind of quality appraisal. This is particularly important as there is evidence, and further research continues in this area - see Cleary P M [8], - that spreadsheets are used as significant tools of business decision-making at even the highest levels of an organisation. However, it is gratifying to see that training organisations such as the ECDL Foundation (European Computer Driving Licence) are providing good quality materials and also tackling the issues of spreadsheet quality. The ECDL is the Europe wide qualification that enables people, without prior knowledge of IT or computer skills, to gain an industry recognised qualification that demonstrates their competence in basic computer skills. The qualification is based around seven modules of which creating and using spreadsheets is one.

## 4. AUDITING

Auditing strategies, methods and accompanying software have an important role to play. There are several commercial tools available: SpACE (Spreadsheet Audit for Customs & Excise), OAK (Operis Analysis Kit), Spreadsheet Detective and Spreadsheet Professional to name a few. An excellent evaluation of several of these products is available in Nixon D [18] and gives a useful starting point for anyone considering the use of such a product. In addition, there is interesting work under way on other tools with differing approaches such as the experimental tool developed at the University of Klagenfurt and used with the open-source spreadsheet system Gnumeric - see Clermont M et al [221]. In addition to the stand-only software audit tools there are the complete spreadsheet audit approaches as used by the typical City of London management consultancies. A good example of such an approach is described in Croll G [25].

# 5. MANAGEMENT

*"The risk of error arising from poor practice in the use of spreadsheets are known to be high, and the incidence of good practice in development using spreadsheets are known to be low, Despite this, users are blissfully unaware of these risks and are using potentially faulty decision support machinery every day to take vital business decisions"* Butler R [6]

There are many, some say too many, issues regarding spreadsheet controls that need to be addressed through organisational policy. There are the issues of choosing and adopting a set of controls to be audited and reviewed at regular intervals. These might include version control of spreadsheets, peer and self-audit approaches to spreadsheet build, team reviews and 'walkthroughs' of spreadsheet models, documentation policies, the use of trustworthy models as templates, and the recording of common errors in a database repository of corporate knowledge. To accomplish even some of these requirements a commitment from higher level management is needed along with a strategy for establishing good practice based upon reasoned selection of standards. Such a strategy may be found in the CobiT (Control Objectives for Information and related Technology) approach.

*"One of the problems reported by researchers and auditors in the field of spreadsheet risks is that of getting and keeping management's attention to the problem. Since 1996, the Information Systems Audit & Control Foundation and the IT Governance Institute have published CobiT\* which brings mainstream IT control issues into the corporate governance arena. This paper illustrates how spreadsheet risk and control issues can be mapped onto the CobiT framework and thus brought to managers' attention in a familiar format"* Butler R [11]

The CobiT approach is based around the Maturity Model for software development capability defined by the Software Engineering Institute. Generally, management uses the model (see fig. 1) to assess the current status of:

- their organisation

- the best practice or the general state of practice in their respective industry

- applicable International standards

And define where the organisation wants to be against these levels.

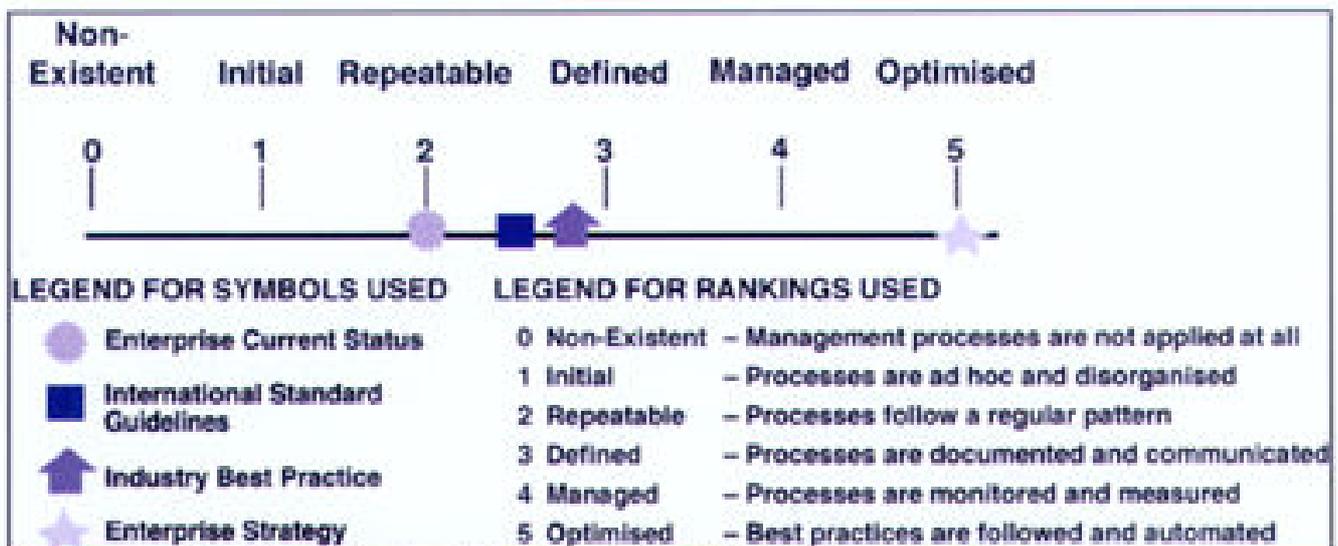

Fig.1 - The CoBIT ® Maturity Model (reprinted from Butler [11]

## 6. CONCLUSION

From the above analysis of past EuSpRIG papers can be seen the comprehensive coverage of ways and means to bring awareness of the issues and to control and manage the problem. Full management of the risks associated with spreadsheets has yet to be achieved and elimination of all errors may remain a dream forever, but as long as a forum exists where practitioners of all kinds can meet, the sooner the topic will become addressed openly within industry.

## 7. REFERENCES

[1] *Spreadsheet Errors: What We Know. What We Think We Can Do.*
Panko R Proceedings of 1st International Symposium on Spreadsheet Risks, EUSPRIG2000, Greenwich, UK, July 2000. http://arxiv.org/abs/0802.3457

[2] *Classification of Spreadsheet Errors*
Rajalingham K, Chadwick D, Knight B: Proceedings of 1st International Symposium on Spreadsheet Risks, EUSPRIG2000, Greenwich, UK, July 2000. http://arxiv.org/abs/0805.4224

[3] *Building Financial Accuracy into Spreadsheets*
Hawker A: Proceedings of 1st International Symposium on Spreadsheet Risks, EUSPRIG2000, Greenwich, UK, July 2000. http://arxiv.org/abs/0805.4219

[4] *Detecting Errors in Spreadsheets*
Ayalew Y, Clermont M, Mittermeir R: Proceedings of 1st International Symposium on Spreadsheet Risks, EUSPRIG2000, Greenwich, UK, July 2000. http://arxiv.org/abs/0805.1740

[5] *A Structured Methodology for Spreadsheet Modelling.*
Rajalingham K, Chadwick D, Knight B; Proceedings of 1st International Symposium on Spreadsheet Risks, EUSPRIG2000, Greenwich, UK, July 2000. http://arxiv.org/abs/0805.4218

[6] *Risk Assessment for Spreadsheet Developments*
Butler R (HM Customs and Excise) Proceedings of 1st International Symposium on Spreadsheet Risks, EUSPRIG2000, Greenwich, UK, July 2000. http://arxiv.org/abs/0805.4236

[7] *Visual Checking Of Spreadsheets*
Chan H Q Chen Y: Proceedings of 1st International Symposium on Spreadsheet Risks, EUSPRIG2000, Greenwich, July 2000. http://arxiv.org/abs/0805.2189

[8] *How Important Are Spreadsheets To Organisations?*
Cleary P M: Proceedings of 1st International Symposium on Spreadsheet Risks, EUSPRIG2000, Greenwich, UK, July 2000.

[9] *Stop the Subversive Spreadsheet*
Chadwick D. Proceedings of 1st International Symposium on Spreadsheet Risks, EUSPRIG2000, Greenwich, July 2000. http://arxiv.org/abs/0712.2594

[10] *Teaching Spreadsheet Development Using Peer Audit and Self-Audit Methods for Reducing Errors*
Chadwick D., Sue R.: Proceedings of 2nd International Symposium on Spreadsheet Risks, EUSPRIG2001, Amsterdam, Holland, July 2001. http://arxiv.org/abs/0801.1514

[11] *Applying the Cobit® Control Framework to Spreadsheet Developments*
Butler R J: Proceedings of 2nd International Symposium on Spreadsheet Risks, EUSPRIG2001, Amsterdam, Holland, July 2001. http://arxiv.org/abs/0801.0609

[12] *Euro Conversion in Spreadsheets*
O'Beirne P: Proceedings of 2[nd] International Symposium on Spreadsheet Risks, EUSPRIG2001, Amsterdam, Holland, July 2001

[13] *Safer Spreadsheets With Model Master*
Paine J: Proceedings of 2[nd] International Symposium on Spreadsheet Risks, EUSPRIG2001, Amsterdam, Holland, July 2001. http://arxiv.org/abs/0801.3690

[14] *An Evaluation of the Quality of a Structured Spreadsheet Development Methodology*
Rajalingham K, Chadwick D, Knight B: Proceedings of 2[nd] International Symposium on Spreadsheet Risks, EUSPRIG2001, Amsterdam, Holland, July 2001.
http://arxiv.org/abs/0801.1516

[15] *New Guidelines for Writing spreadsheets*
Raffensperger J: Proceedings of 2[nd] International Symposium on Spreadsheet Risks, EUSPRIG2001, Amsterdam, Holland, July 2001.

[16] *Assurance By Control Around Is A Visible Alternative To The Traditional Approach*
Ettema H, Janssen P, de Swart J: Proceedings of 2[nd] International Symposium on Spreadsheet Risks, EUSPRIG2001, Amsterdam, Holland, July 2001. http://arxiv.org/abs/0801.4775

[17] *A Real alternative to spreadsheets*
Knight D,: Proceedings of 2[nd] International Symposium on Spreadsheet Risks, EUSPRIG2001, Amsterdam, Holland, July 2001

[18] *Spreadsheet Auditing*
Nixon D, O'Hara M: Proceedings of 2[nd] International Symposium on Spreadsheet Risks, EUSPRIG2001, Amsterdam, Holland, July 2001

[19] *Training Gamble Leads To Corporate Grumble*
Chadwick D: Proceedings of 3[rd] International Symposium on Spreadsheet Risks –the Hidden Corporate Gamble, EUSPRIG 2002, Cardiff, Wales, UK, July 2002

[20] *Interpretation as a Factor in Understanding Flawed Spreadsheets*
Banks D., Monday A.: Proceedings of 3[rd] International Symposium on Spreadsheet Risks -the Hidden Corporate Gamble, EUSPRIG 2002, Cardiff, Wales, UK, July 2002.
http://arxiv.org/abs/0801.1856

[21] *Spreadsheet Engineering: A Research Framework*
Grossman T.: Proceedings of 3[rd] International Symposium on Spreadsheet Risks - the Hidden Corporate Gamble, EUSPRIG 2002, Cardiff, Wales, UK, July 2002.
http://arxiv.org/abs/0711.0538

[22]*A Spreadsheet Auditing Tool Evaluated In An Industrial Context*
Clermont M., Hanin C., Mittermeir R.: Proceedings of 3d International Symposium on Spreadsheet Risks - the Hidden Corporate Gamble, EUSPRIG 2002, Cardiff, Wales, UK, July 2002. http://arxiv.org/abs/0805.1741

[23] *The Subversive Spreadsheet*

Chadwick D., Knight B.: Proceedings of 3d International Symposium on Spreadsheet Risks - the Hidden Corporate Gamble, EUSPRIG 2002, Cardiff, Wales, UK, July 2002.

[24] *Losing At Spreadsheet Roulette*
Butler R.: Proceedings of 3rd International Symposium on Spreadsheet Risks - the Hidden Corporate Gamble, EUSPRIG 2002, Cardiff, Wales, UK, July 2002

[25] *A Typical Spreadsheet Audit Approach*
Croll G.: Proceedings of 3d International Symposium on Spreadsheet Risks - the Hidden Corporate Gamble, EUSPRIG 2002, Cardiff, Wales, UK, July 2002.
http://arxiv.org/abs/0712.2591